\documentclass[a4paper,12pt]{report}
 \usepackage{graphicx,epsf}
 \usepackage{bm}
 \usepackage{amssymb}

\title{Observation of inertial energy cascade\\
 in interplanetary space plasma}
\author{Luca Sorriso-Valvo$^1$, Raffaele Marino$^2$, Vincenzo Carbone$^2$, \\
Fabio Lepreti$^2$, Pierluigi Veltri$^2$, Alain Noullez$^3$, \\
Roberto Bruno$^4$, Bruno Bavassano$^4$, Ermanno Pietropaolo$^5$ \\
\\
\\
$^1$ Licryl Regional Laboratory - INFM/CNR, Ponte P. Bucci, \\
Cubo 33C, 87036 Rende (CS), Italy\\
$^2$ Dipartimento di Fisica, Universit\`a della Calabria, and CNISM,\\
Ponte P. Bucci Cubo 31C, 87036 Rende (CS), Italy \\
$^3$ Observatoire de la C\^ote d'Azur, \\
Boulevard de l'Observatoire, 06304 Nice, France \\
$^4$ IFSI/INAF, Via Fosso del Cavaliere, 00133 Roma, Italy \\
$^5$ Dipartimento di Fisica, Universit\`a dell'Aquila, \\
Via Vetoio, 67010, L'Aquila, Italy}

\begin{document}
\maketitle

\begin{abstract}
We show in this article direct evidence for the presence of an
inertial energy cascade, the most characteristic signature of
hydromagnetic turbulence (MHD), in the solar wind as observed by the
Ulysses spacecraft.  After a brief rederivation of the equivalent
of Yaglom's law for MHD~turbulence, we show that a linear relation is
indeed observed for the scaling of mixed third order structure
functions involving Els\"asser variables. This experimental result, 
confirming the prescription stemming from a theorem for MHD turbulence, firmly
establishes the turbulent character of low-frequency velocity and magnetic
field fluctuations in the solar wind plasma.
\end{abstract}

\newpage

\noindent 
Space flights have shown that the interplanetary medium is
permeated by a supersonic, highly turbulent plasma flowing out from the solar
corona, the so called solar wind~\cite{marsch,noi}.  
The turbulent character of the flow, at frequencies below the ion gyrofrequency~$f_{ci}\simeq 1$Hz, 
has been invoked since the first Mariner
mission~\cite{coleman}.  In fact, velocity and magnetic
fluctuations power spectra are close to the
Kolmogorov's~-5/3~law~\cite{noi,frisch}.  However, even if fields fluctuations 
are usually considered within the framework of magnetohydrodynamic (MHD)
turbulence~\cite{noi}, a firm established proof of the existence
of an energy cascade, namely the main characteristic of
turbulence, remains a conjecture so far~\cite{DMV}. 
This apparent lack could be fulfilled through the evidence for
the existence of the only exact and nontrivial result of turbulence~\cite{frisch}, 
that is a relation between the third order moment of the
longitudinal increments of the fields and the separation~\cite{kolmogorov}.
This observation would firmly put low frequency solar wind
fluctuations within the framework of MHD turbulence.  
The importance of such question stands beyond the
understanding of the basic physics of solar wind turbulence.
For example, it is well known that turbulence is one of the main obstacles
to the confinement of plasmas in the fusion devices~\cite{fusion,fusiontur}.
The understanding of interplanetary turbulence and its effects on energetic particle
transport is of great importance
also for Space Weather research~\cite{kumar}, which is a relevant
issue for spacecrafts and communication satellites operations,
and for the security of human beings.
Finally, more theoretical problems are concerned, such as the puzzle of solar 
coronal heating due to the turbulent flux toward small scales~\cite{corona}.

Incompressible MHD~equations are more complicated than the standard neutral
fluid mechanics equations because the velocity of the charged fluid is coupled
with the magnetic field generated by the motion of the fluid itself.
However, written in terms of the Els\"asser variables defined as ${\bm z}^\pm =
{\bm v} \pm (4\pi\rho)^{-1/2}{\bm b}$ (${\bm v}$ and~${\bm b}$ are the velocity
and magnetic field respectively and $\rho$~the mass density), they have the
same structure as the Navier-Stokes equations~\cite{DMV}
 \begin{equation}
 \partial_t {\bm z}^\pm+{\bm z}^\mp \cdot {\bm \nabla}{\bm z}^\pm
 = -{\bm \nabla} P/\rho
 +\left({\nu+\kappa \over 2}\right) \nabla^2 {\bm z}^\pm
 +\left({\nu-\kappa \over 2}\right) \nabla^2 {\bm z}^\mp \ ,
 \label{e:mhd}
 \end{equation}
where~$P$ is the total hydromagnetic pressure, while $\nu$ is the viscosity and
$\kappa$ the magnetic diffusivity.  In particular, the nonlinear term appears
as ${\bm z}^\mp \cdot {\bm \nabla}{\bm z}^\pm$, suggesting the form of a
transport process, in which Alfv\'enic MHD~fluctuations~${\bm z}^\pm$
propagating along the background magnetic field are transported by
fluctuations~${\bm z}^\mp$ propagating in the opposite direction.  This
transport is {\em active\/} as ${\bm z}^\pm$ and ${\bm z}^\mp$ are clearly not
independent.  Still, following the same procedure as in~\cite{antonia,danaila},
and assuming local homogeneity, a relation similar to the Yaglom equation for
the transport of a {\em passive\/} quantity~\cite{monin} can be obtained in the
stationary state
 \begin{eqnarray}
 \partial_\parallel Y^\pm(r) &=&
 -{4 \over 3} \,\epsilon^\pm
 + 2\nu\, \nabla^2 \left\langle |\Delta {\bm z}^\pm|^2 \right\rangle
 \nonumber\\ &&
 -2\left\langle \Delta {\bm z}^\pm \cdot
 ({\bm \nabla}+{\bm \nabla}')\Delta P/\rho \right\rangle
 +\left\langle {\bm z}^\mp \cdot
 ({\bm \nabla}+{\bm \nabla}')|\Delta {\bm z}^\pm|^2 \right\rangle \ .
 \label{yaglom}
 \end{eqnarray}
Here, $\Delta {\bm z}^\pm \equiv {\bm z}^\pm({\bm x}')-{\bm z}^\pm({\bm
x})$~are the (vector) increments of the fluctuations between two points~${\bm
x}$ and~${\bm x}' \equiv {\bm x}+{\bm r}$, ${\bm \nabla}$ and~${\bm \nabla}'$
are the gradients at the corresponding two points, $\partial_\parallel$ is the
longitudinal derivative along the separation~${\bm r}$, while $Y^\pm(r)$~are
the mixed third order structure function~$\langle |\Delta {\bm z}^\pm|^2\,
\Delta z^\mp_\parallel \rangle$ and $\epsilon^{\pm} \equiv \nu\, \left\langle
|{\bm \nabla}{\bm z}^\pm|^2 \right\rangle \stackrel{\rm hom}{=} 3\nu\,
\left\langle |\partial_\parallel\, {\bm z}^\pm|^2 \right\rangle$ are the
pseudo-energy average dissipation rates, namely the dissipation rates of
both~$\left\langle |{\bm z}^\pm|^2 \right\rangle /2$ respectively.  Finally,
$\Delta P$~represent the increment of the total pressure fluctuations and the
kinematic viscosity~$\nu$ is here assumed to be equal to the magnetic
diffusivity~$\kappa$ (this last assumption is in fact not necessary if we
concentrate on the inertial range, as we will do from now on).

The last term on the~r.h.s. of equation~(\ref{yaglom}) is related to
large-scale inhomogeneities and disappears if the flow is globally homogeneous.
Also, assuming local isotropy, the term containing pressure correlation
vanishes, so that after longitudinal integration of~(\ref{yaglom}) and in the
inertial range of MHD~turbulence (i.e. when~$\nu \to 0$), a linear scaling law
 \begin{equation}
 Y^\pm(r) = -{4 \over 3} \,\epsilon^\pm\, r
 \label{linear}
 \end{equation}
is obtained, characterizing a turbulent cascade with a well-defined finite
energy flux~$\epsilon^{\pm}$.  An alternative derivation of this result using
correlators instead of structure functions was also obtained in~\cite{PoPo}, 
and observed in numerical simulations~\cite{sorriso02}.
When neutral fluid turbulence is considered, equation~(\ref{linear})
becomes~\cite{antonia}~$\langle |\Delta {\bm v}|^2\, \Delta v_\parallel \rangle
= -4/3 \,\epsilon\, r$ ($\epsilon$~being the average kinetic energy dissipation
rate), from which Kolmogorov's~-4/5~law for the longitudinal third order
structure function can be recovered if there is full isotropy as~$\langle
(\Delta v_\parallel)^3 \rangle = -4/5 \,\epsilon\, r$.

In this work, we show that relation~(\ref{linear}) is indeed
satisfied in some periods within solar wind turbulence. In order
to avoid variations of the solar activity and ecliptic
disturbances (like slow wind sources, Coronal Mass Ejections,
ecliptic current sheet, and so on), we use high speed polar wind
data measured by the Ulysses spacecraft~\cite{uly1,uly2}.
In particular, we analyse here the first seven months of~1996,
when the heliocentric distance slowly increased from 3~AU to 4~AU, 
while the heliolatitude decreased from about~$55^{\circ}$ to~$30^{\circ}$. 
The fields components are given in the $RTN$~reference frame, where
$R$~(radial) indicates the sun-spacecraft direction, centered on
the spacecraft and pointing out of the sun, $N$~(normal) lies in
the plane containing the radial direction and the sun rotation
axis, while $T$~completes the right-handed reference frame. Note
that, since the wind speed in the spacecraft frame is much larger
than the typical velocity fluctuations, and is nearly aligned
with the $R$~radial direction, time fluctuations are in fact
spatial fluctuations with time and space scales ($\tau$ and~$r$
respectively) related through the Taylor hypothesis, so that $r =
-\langle v_R \rangle \tau$ (note the {\em reversed\/} sign). From
the 8~minutes averaged time series~${\bm z}^\pm(t)$, we compute
the time increments~$\Delta {\bm z}^\pm(\tau;t) = {\bm
z}^\pm(t+\tau)-{\bm z}^\pm(t)$, and obtain the mixed third order
structure function~$Y^\pm(-\langle v_R \rangle_t\, \tau) =
\left\langle |\Delta {\bm z}^\pm(\tau;t)|^2\, \Delta
z^\mp_R(\tau;t) \right\rangle_t$ using moving averages~$\langle
\bullet \rangle_t$ on the time~$t$ over periods spanning
around~10~days, during which the fields can be considered
stationary.

\begin{figure}
\epsfxsize=20cm \centerline{\epsffile{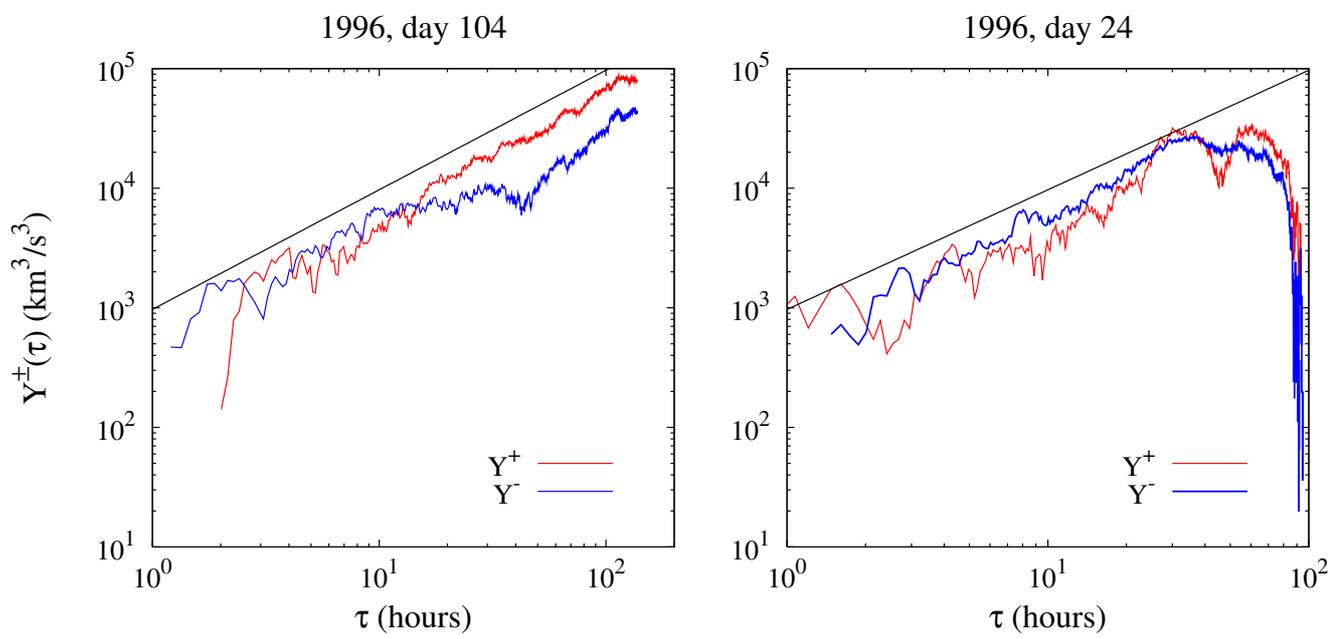}} \caption{The
scaling behaviour of both~$Y^\pm(\tau)$ as a function of the
scale~$\tau$ for two different periods we examined. The full
black line shows a linear scaling law to guide the eye.}
\label{fig_scaling}
\end{figure}

\begin{figure}
\epsfxsize=18cm
\centerline{\epsffile{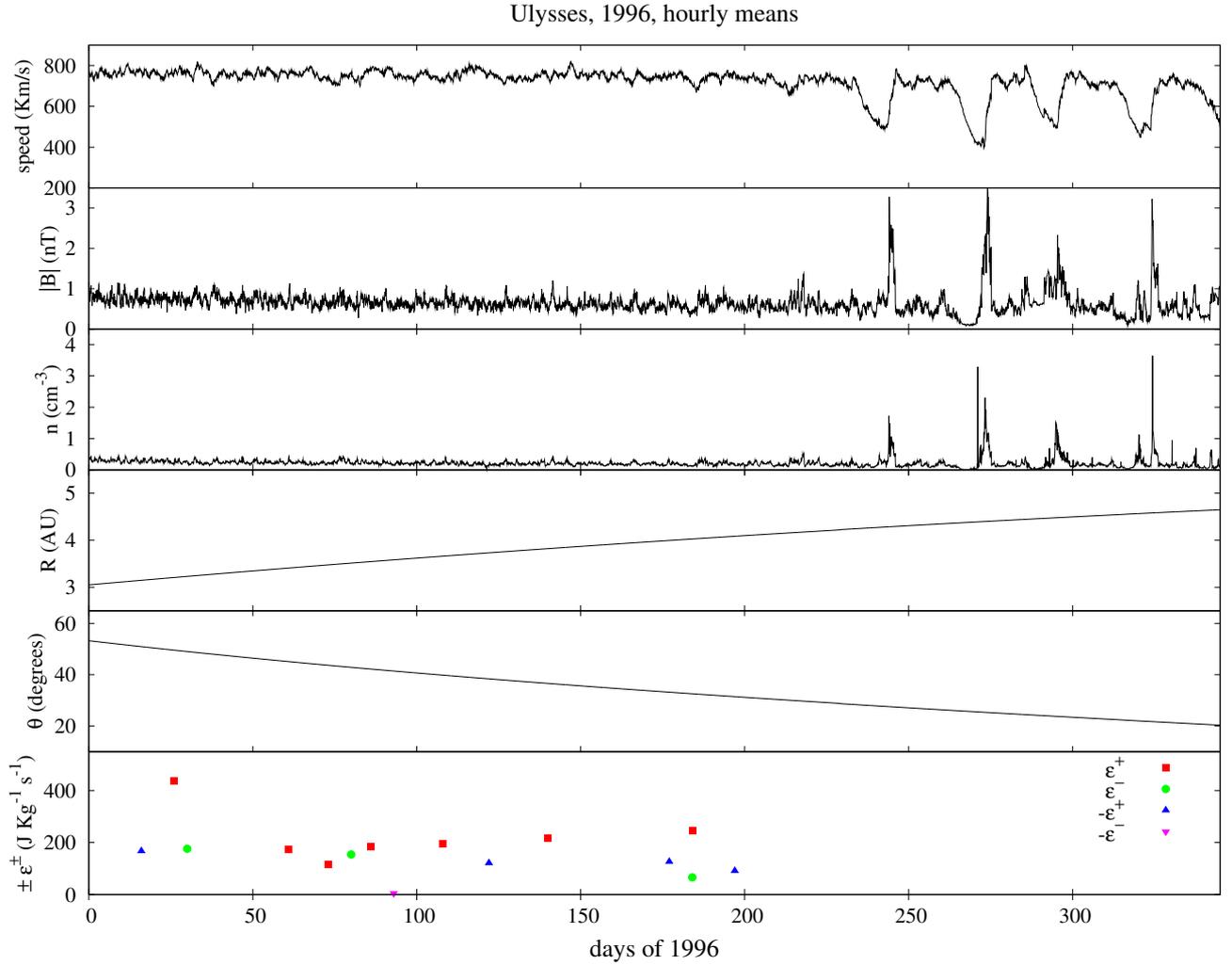}}
\caption{Hourly averaged quantities are represented as a function of the flight
time of Ulysses.  The top panels represent respectively the solar
wind speed, the magnitude of the magnetic field, the particle density, the
distance from the sun and the heliolatitude angle.  In the bottom panel
 the values of~$\epsilon^\pm$, calculated through a fit with the
function~(\ref{linear}) during the periods where a clear linear scaling
exists, are reported.}
\label{fig_epsilon}
\end{figure}

A linear scaling~$Y^\pm(\tau) = 4/3 \,\epsilon^\pm\, \langle v_R
\rangle_t\, \tau$ is indeed observed in a significant fraction of
the periods we examined, with an inertial range spanning as much
as two decades, indicating the existence of
a well defined inertial energy cascade range in plasma
turbulence. In fact, solar wind inertial ranges can even be
larger than the ones reported for laboratory fluid
flows~\cite{danaila}, showing the robustness of this result.
This is the first experimental validation of the turbulence MHD theorem
discussed above.
Figure~\ref{fig_scaling} shows an example of scaling and the
extension of the inertial range, for both~$Y^\pm(\tau)$.  The
linear scaling law generally extends from a few minutes to one
day or more. This happens in about 20~periods of a few days in
the 7~months considered. Several other periods are found in which
the scaling range is considerably reduced. In particular, the
sign of $Y^\pm(\tau)$ is observed to be either positive or
negative. Since pseudo-energies dissipation rates are positive
defined, a positive sign for~$Y^\pm(\tau)$ (negative
for~$Y^\pm(r)$) indicates a (standard) forward cascade with
pseudo-energies flowing towards the small scales to be
dissipated.  On the contrary, a negative~$Y^\pm(\tau)$ is the
signature of an {\em inverse\/} cascade where the energy flux
is being transferred on average toward larger scales. 
Figure~\ref{fig_epsilon} shows the location of the most evident scaling intervals, 
together with the values of the flux rate~$\epsilon^\pm$ estimated 
through a fit of the scaling law (\ref{linear}), typically of the order 
of a few hundreds in~${\rm J}\,{\rm kg}^{-1}\,{\rm s}^{-1}$. 
For comparison, values found for ordinary turbulent fluids are 
$1\div 50$ ${\rm J}\,{\rm kg}^{-1}\,{\rm s}^{-1}$~\cite{meneveau}.

\begin{figure}
\epsfxsize=20cm
\centerline{\epsffile{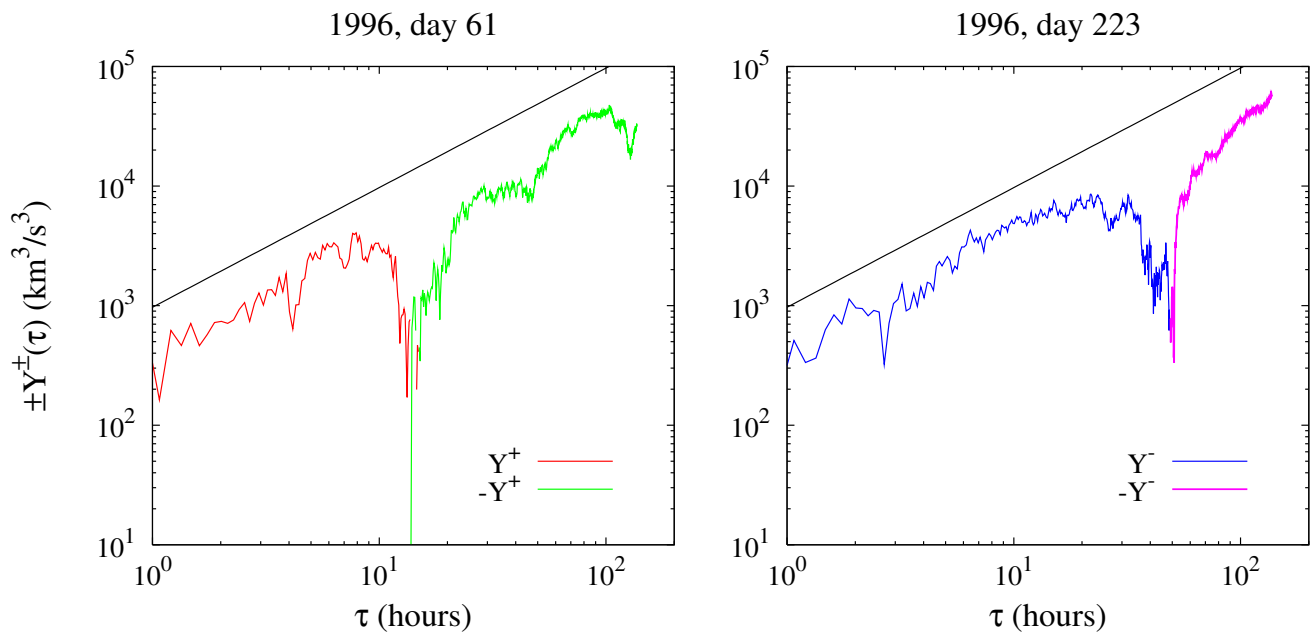}}
\caption{The scaling behaviour of~$Y^\pm(\tau)$ as a function of the time
scale~$\tau$ for two different periods we examined.  Different colours of the
curves refer to positive and negative values of the mixed structure
functions~$Y^\pm(\tau)$ and thus of~$\epsilon^\pm$.  The full black line
correspond to a linear scaling law.}
\label{fig_pos-neg}
\end{figure}

It is worth noting that, in a large fraction of cases,
both~$Y^\pm(\tau)$ switch from positive to negative  linear
scaling (or viceversa) within the same time period  when going
from small to large scales (see Figure~\ref{fig_pos-neg}). The
occurrence of both kind of cascades within the same flow is not
so uncommon within hydrodynamic turbulence. This phenomenon has
been attributed to some large scale instability, as observed for
example in geophysical flows or when the flow is affected by a
strong rotation~\cite{moisy}. In the case of solar wind plasma a
possible explanation for the inverse cascade could be the
enhanced intensity of the background magnetic field. This would
make the turbulence mainly bidimensional allowing for an inverse
cascade as observed in numerical simulations~\cite{dued}. It
should also be noticed that in most of the cases the time scale at
which the cascade reverses its sign is of the order of 1 day.
This scale roughly indicates where the small scale Alfv\'enic
correlations between velocity and magnetic field are 
lost \cite{bruno85, goldstein95}. This could
mean that the nature of the fluctuations changes across the break.
However, these particular aspects still deserve to be adequately
considered within the solar wind context.

At this point, the question is why the scaling is not observed
all the time within the solar wind. As already stated, equation~(\ref{yaglom}) 
reduces to the linear law~(\ref{linear}) only when local homogeneity,
incompressiblity and isotropy conditions are satisfied. In
general, solar wind inhomogeneities play a major role at large
scales so that local homogeneity is generally fulfilled within
the range of interest. Regarding incompressibility, it has been shown that compressive
phenomena mainly affect shocked regions and dynamical interaction
regions like stream-stream interface \cite{marsch,noi}. However,
the time interval we analyze, because of Ulysses high latitude
location, is not affected by these compressive phenomena
\cite{gosling1995}. On the other hand, it has also been shown
\cite{noi} that magnetic field compressibility increases mainly
at very small scales within the fast wind regime.  It follows
that the incompressibility assumption can be considered valid to
a large extent for the analyzed interval and at intermediate
scales.The large scale anisotropy, mainly due to the average magnetic
field, is only partially lost at smaller scales, and a residual
anisotropy is always present~\cite{tim,epl}, generally breaking
the local isotropy assumption. Thus, while inhomogeneity, compressibility and anisotropy could
all be responsible for the loss of linear scaling, anisotropy
probably is the main candidate within high latitude regions of
the solar wind.

It is important to note that the presence of a Yaglom-like law
that involves the third order mixed moment, is more general than
the phenomenology usually involving the second order moment.
Indeed, while the Yaglom MHD relation~(\ref{yaglom}) involves only
differences along the parallel direction, that are in fact the
only quantities accessible from single satellite measurements,
phenomenological arguments involve the full spatial dependence of
vector fields that cannot be directly measured yet. This means
that our result is compatible with both Kolmogorov and
Iroshnikov-Kraichnan type cascade~\cite{DMV,noi}, and does not
help in discriminating between these
phenomenologies~\cite{debate1,debate2}.

In conclusion, we observed, for the first time in the solar wind, the only
natural plasma directly accessible, evidence of Yaglom MHD
scaling law indicating the existence of a local energy cascade in
hydromagnetic turbulence. The scaling holds in a number of
relatively long periods of about 10~days, and also provides the
first estimation of the pseudo-energy dissipation rate. Although
our data might not fully satisfy requirements of homogeneity,
incompressibility and isotropy everywhere, the observed linear
scaling extends on a wide range of scales and appears very robust.  The
unexpected existence of the scaling law in anisotropic, weakly
compressible and inhomogeneous turbulence still needs to be fully
understood. Our result estabilishes a firm point within solar wind phenomenology, 
and, more generally, provides a better knowledge of plasma turbulence, 
carrying along a wide range of practical implications on both laboratory 
fusion plasmas and space physics.

\noindent {\bf Acknowledgments} The use of data of the plasma analyzer (principal investigator D. J.
McComas, Southwest Research Institute, San Antonio, Texas, USA) and of the
magnetometers (principal investigator A. Balogh, The Blackett Laboratory,
Imperial College, London, UK) aboard the Ulysses spacecraft is gratefully
acknowledged. The data have been made available through the World Data
Center A for Rockets and Satellites (NASA/GSFC, Greenbelt, Maryland, USA).

\newpage

\end{document}